# Visualization tools for uncertainty and sensitivity analyses on thermal-hydraulic transients

Anne-Laure Popelin[1*] and Bertrand Iooss[1]

[1]EDF R&D, Industrial Risk Management Department, 6 Quai Watier 78401 Chatou, France
*Corresponding Author, E-mail: anne-laure.popelin@edf.fr

In nuclear engineering studies, uncertainty and sensitivity analyses of simulation computer codes can be faced to the complexity of the input and/or the output variables. If these variables represent a transient or a spatial phenomenon, the difficulty is to provide tool adapted to their functional nature. In this paper, we describe useful visualization tools in the context of uncertainty analysis of model transient outputs. Our application involves thermal-hydraulic computations for safety studies of nuclear pressurized water reactors.

## I. Introduction

This work is part of current research concerning engineering studies of pressurized water reactor. EDF R&D and its partners develop generic probabilistic approaches for the uncertainty management of computer code used in safety analyses[1]. One of the main difficulties in uncertainty and sensitivity analyses is to deal with thermal-hydraulic computer code[2]. Indeed, most of mathematical tools are adapted to scalar input and output variables, while the outputs of thermal-hydraulic models represent time-dependent state variable (temperature, pressure, thermal exchange coefficient, etc.).

As an industrial example, we will consider the Benchmark for Uncertainty Analysis in Best-Estimate Modelling for Design, Operation and Safety Analysis of Light Water Reactors[3] proposed by the Nuclear Energy Agency of the Organization for Economic Co-operation and Development (OCDE/NEA). One of the study-cases corresponds to the calculation of LOFT L2.5 experiment, which simulated a large-break loss of primary coolant accident. It has been implemented on the French thermal-hydraulic computer code CATHARE2, developed at the Commissariat à l'Energie Atomique (CEA). One phase of the benchmark consists in applying on this study-case the so-called BEMUSE (Best Estimate Methods, Uncertainty and Sensitivity Evaluation) program in order to tests various uncertainty and sensitivity analysis methods[3].

Figure 1 illustrates the BEMUSE data, 100 Monte Carlo simulations (by randomly varying around 50 uncertain inputs of the LOFT L2.5 scenario), given by CATHARE2, of the cladding temperature in function of time. When looking at the overall behavior of this large number of curves, the main questions which arise are the following:

1. What is the average curve?
2. Can we define some confidence interval curves containing most of the curves?
3. Can we detect some abnormal curves, in the sense of a strong difference from the majority of the curves (as outliers for scalar variables)?
4. Are there some clusters which correspond to different behavior of the physical behavior of the output?
5. …

Question 4 has been treated in a previous work by Auder et al.[2]. In this work, we consider the other questions by the way of some functional data analysis tools.

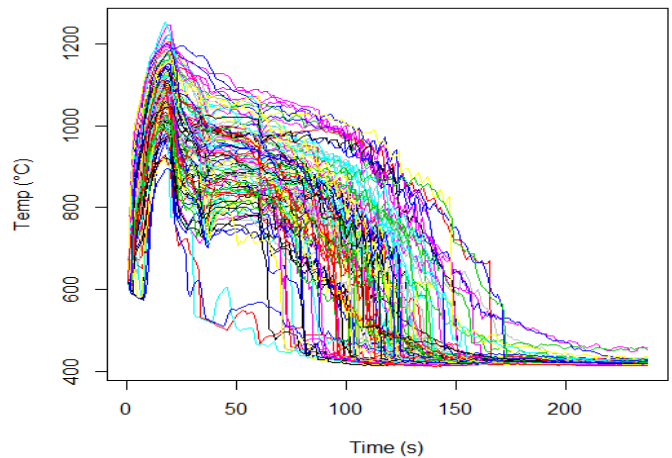

**Figure 1: Visualization of the BEMUSE data: 100 temporal curves of the cladding temperature of the fuel rods. Each curve comes from the output of one computer code (CATHARE2) run. Source: CEA.**

A first problem consists in visualizing the uncertainty of a time-dependent variable, which is called in statistics a functional variable. Regarding their capacity to summarize rich and complex information, visualization tools are very important in statistical studies. Several types of visualization exist, depending on the information that have to be emphasized and the dimension of the involved problems.

In this context, this article deals with studying methodologies meeting two main objectives: detecting some central tendency behavior and outliers; and linking these particular shapes of transients to the input values or combinations of inputs that have induced them.

Two classes of methods have been identified as potential useful tools: a classical way to handle functional variables in statistics is to reduce their dimension via projection or regression techniques[4]; another one is to consider the concept of *band depth*[5]. The first part of this article presents two methods introduced by Hyndman and Shang[6], based on dimension reduction. The second part is dedicated to the functional boxplot of Sun and Genton[7], which implies depth of band concept.

## II. Methods based on dimension reduction

This section presents two methods from an article by Hyndman and Shang[6] for the construction of "functional boxplots", which are both based on a dimension reduction.

Classical boxplot for scalar variables is a very common statistical tool that allows summarizing the main information of a data sample: median, first and third quartiles, and an interquartile-based interval which define the limit of non-outliers data. First step to build such boxplot is to rank data thanks to a statistical order; such order has first to be defined for functional data, which has lead to numerous research works in the literature.

**1. Some background on dimension reduction**

The goal of dimension reduction is to represent the source data into a new space with reduced dimensions, where it will be easier to study. The transformation should keep enough interesting information about the source data, while allowing simplifying the analysis. There are many methods for dimension reduction. Here, we focus only on the principal component analysis (PCA) in its classical and robust variants.

Once in the new space, we can use different methods to estimate the quantiles and outliers. Stages for the size reduction stages and estimates depths are independent from each other. Thus, it is possible to combine several methods, for the reduction itself on one side, and for the calculation of depths on the other side.

In classical linear PCA method, the aim is to find the orthogonal axis on which the projection of the data matrix has a maximized variance. The disadvantage of this method is the lack of robustness regarding extreme values. A variant of this criterion is to use another variable to be maximized. For instance, the Median Absolute Variance criteria offers to maximize the median of the absolute differences between the samples and the median, which are more robust to the extreme values than the mean value.

Once the functional variable has been transformed to a fewer component space (typically two, but this dimension can be increased), the objective is to estimate quantiles in this space, in order to detect outliers.

Remark on explained variance:

The variance explained is an important factor in assessing the performance of the PCA component. For standard PCA, obtaining the variance explained by $k$ components is a simple calculation of the percentage of $k$ eigenvalues of the covariance matrix associated with these components $k$.

We note that using two main components, the explained variance for BEMUSE data is 86.57%. Explained variance increases with the number of principal components up to 95%, with 6 main components for data sets.

## 2. Highest Density Regions method

The principle of this method is to assimilate observations in the space of principal components to the realizations of a random vector with density $f$. By calculating an estimate of the density $f$, the quantiles can then be computed.

A Gaussian smoothing kernel is used in the paper of Hyndman & Shang[6]:

$$\hat{f}(X) = \frac{1}{n} \cdot \sum_{i=1}^{n} K_H(X - X_i)$$

with $K_H(X) = |H|^{-\frac{1}{2}} \cdot K\left(H^{-\frac{1}{2}} \cdot X\right)$,

$K(X) = \frac{1}{2\pi} \cdot \exp\left(-\frac{1}{2}\langle X, X \rangle\right)$ the "standard" Gaussian kernel and the $H$ matrix containing the smoothing parameters. Depending on this matrix (diagonal or not), some preferential smoothing directions can be chosen.

Once the estimate of $f$ is obtained, the higher density regions are considered as deeper data. Thus, on figure 2 the 50% quantile is represented with dark grey, and lighter grey zone is the 95% quantile zone. Points outside this zone are considered as outliers.

The global mode $X_{\max} = \arg\max \hat{f}(X)$ is considered as the deepest curve. Note that this point does not necessarily match with a real curve in the data sample, since f is defined in every point of the component domain.

This method needs to have an a priori on the number of outliers. In some cases, several outliers can be in the same region, and thus wrongly create a high density region, as shown on figure 3.

When returning in the initial space, because of the loss of information from dimension reduction step, the dark and light grey zones do not represent strictly the same information.

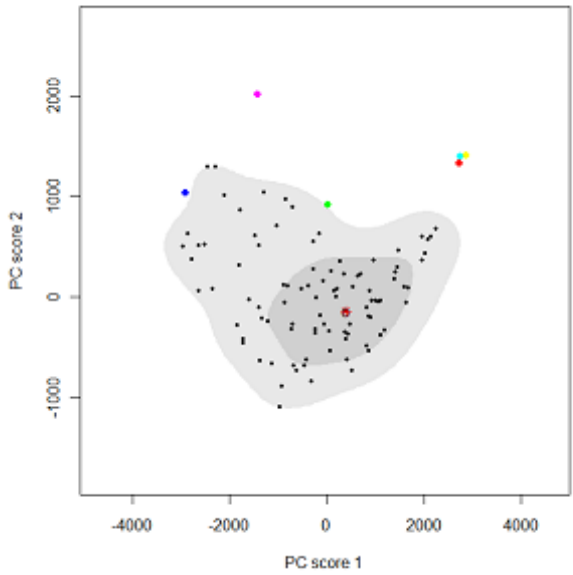

**Figure 2: Visualization of the density estimator on the BEMUSE study case. Dark grey zone envelops 50% of the distribution. Light grey represents 95% quantile zone.**

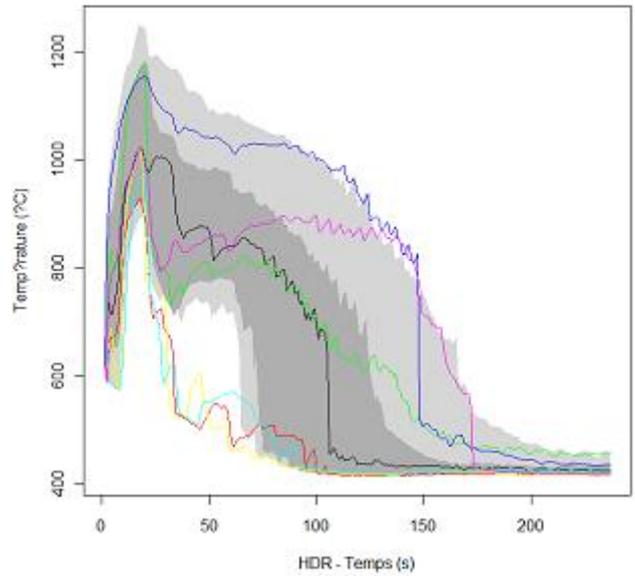

**Figure 4: BEMUSE study-case: Visualization of functional quantiles and outliers, back in the physical space.**

### 3. Bagplot method

The bagplot method was proposed by Rousseeuw et al.[8] for the analysis of bivariate data. This is a generalization of the classical boxplot, an example of which is shown in figure 5. The dark blue area contains 50% of the data in the "center" of the distribution. The light blue area contains data that are "less central" without being considered outliers. Finally, four points above are detected as outliers.

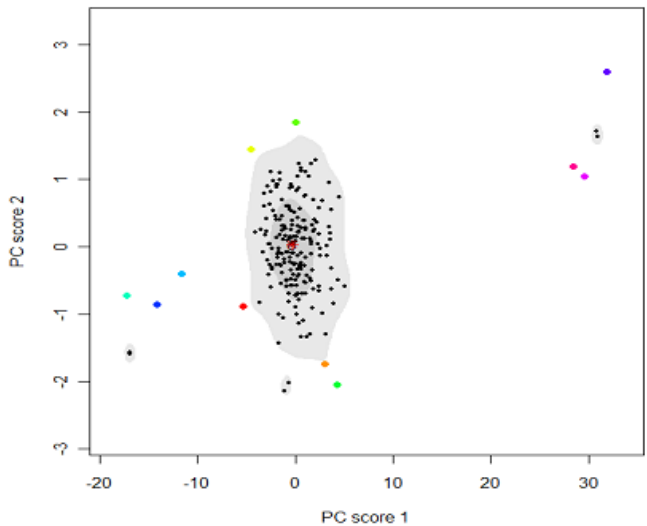

**Figure 3: Visualization of a density estimator on another example. Some outliers being very close from each other: high density regions may be wrong in this case.**

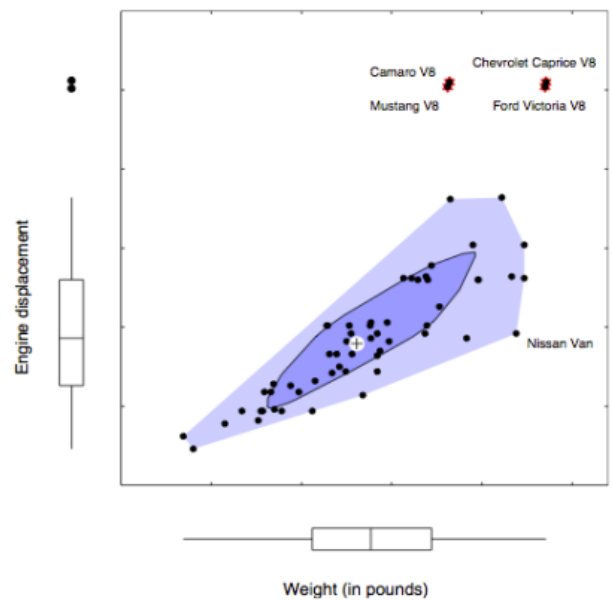

**Figure 4: Example of bagplot (Source: Rousseeuw et al.[8]).**

The construction of bagplot is based on the notion of depth of Tukey[9]. The depth of Tukey at one point $\theta$, relatively to a set of points noted Z, is defined by:

$$L_{\text{depth}}(\theta, Z) = \min \text{card}\{P \cap Z \,; P \in DP(d(\theta))\}$$

where $DP(\theta)$ is a closed half-plane whose boundary is a line containing $\theta$ (Figure 5). The Tukey depth is defined for all points in the plane, not only the experimental data.

We define the "median" as the point at which the depth is higher. The definition of Tukey depth can easily be generalized to higher dimensions, but the calculation becomes extremely expensive.

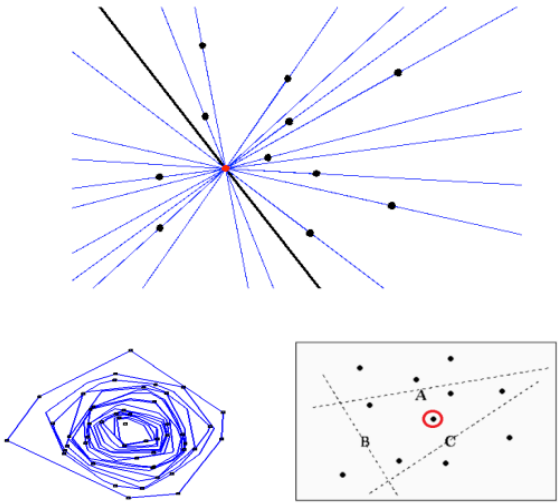

**Figure 5: Illustration of the depth of Tukey.**

I in the following, four steps are detailed for the construction of a bagplot.

1) <u>Ranking points according to their depth:</u> sort the points and draw the iso-depth contours. Many algorithms have been proposed, such as algorithm FDC proposed by Johnson et al.[10]. This classification allows building the central region containing 50% of points that have the highest depth.

2) <u>Defining the median point:</u> find the median which is the point with the highest depth (this is not necessarily an experimental point). There are many algorithms to solve this problem, for example: HALFMED algorithm proposed by Rousseeuw and Ruts[11].

3) <u>Detecting outliers:</u> It is an "expansion in depth" of the central region. Points whose depth is less than

$$Pf_{\text{limite}} = P_{\text{médian}} - |P_{\text{médian}} - Pf_{\text{bag}}| * \text{coef}$$

are considered as outliers. Regarding the coefficient, Rousseeuw et al.[8] proposed the value 3 while Hyndman and Shang[6] suggest 2.57 because this value retains 99% of the points in the case of a Gaussian distribution. A confidence interval around the median point is constructed by using the bootstrap method proposed by Febrero et al.[12].

4) <u>Representation of points in the functional space:</u> The envelope of the curves contained in the central region is colored in dark gray. The confidence interval for the median point is presented by dotted lines. The outlier curves are drawn in different colors. Envelope of the other curves (not outliers) is colored in light gray (see Figures 6 and 7 for the application of this tool to the BEMUSE data).

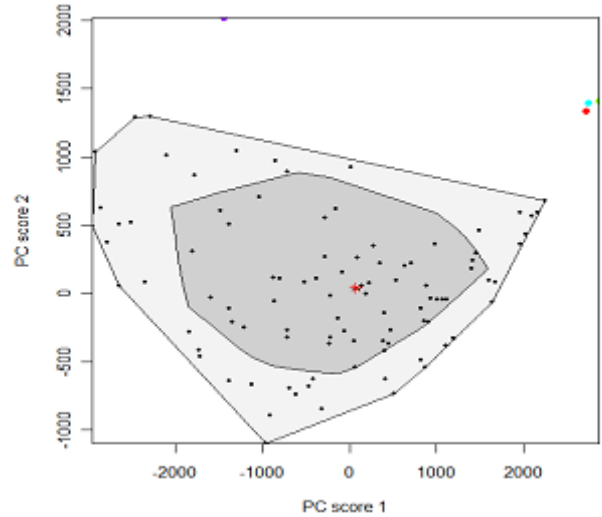

**Figure 6: Bagplot of BEMUSE study case.**

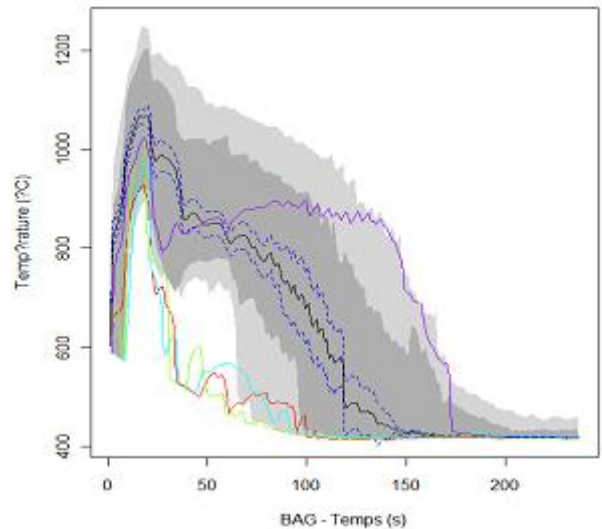

**Figure 7: Bagplot results in the functional space for BEMUSE study-case.**

Remarks on bagplot :

1) Unimodality
The bagplot implies unimodality. The figure below presents a difficult case to deal with this method. We see that the median is always detected in the center of the middle zone, while it is not always what we would get the most representative curve (Figure 8). This is the big difference between bagplot and HDR plot. HDR plot may well separate the different modes of distribution, but this is not the case with bagplot.

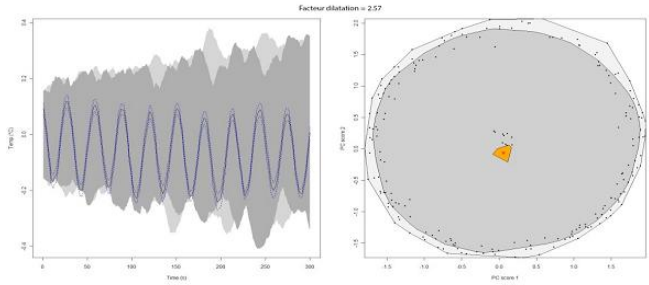

**Figure 8: Example of a non-unimodal problem, where bagplot is not relevant.**

2) Generalization to high dimensions
Theoretically, bagplot can be generalized for large dimensions. There are powerful algorithms to determine the median point in higher dimensions, such as DEEPLOC algorithm proposed by Rousseeuw and Struyf[13].
However, tracing the contours of iso-depth in dimension greater than 3 is a difficult problem. Nevertheless, Chen et al.[14] propose an algorithm using the Monte Carlo method.

## III. Methods using band depth concept

This section presents a method for the visualization and analysis of functional data developed by Sun and Genton[7]. This method is based on the notion of depth band that classifies a sample of curves.
To generalize the scheduling of a statistical sample, we introduced different versions of "depth of data." A "depth" is associated with each element of the sample, which allows classifying and thus finding the concepts of median and outliers. For functional data, Lopez-Pintado and Romo[5] introduced a notion of band depth. This allows classifying a set of curves and thus defining functional quantiles, to identify the most central (median) curves and outliers curves
Each curve is associated with a real that is the band depth. Specifically, from a sample of curves $(y_1(t), \ldots, y_n(t))$, band depth will allow to obtain a ranked sample:
$(y_{[1]}(t), \ldots, y_{[n]}(t))$ where $y_{[1]}(t)$ is the deepest curve and $y_{[n]}(t)$ is the least one. The $y_{[1]}(t)$ curve plays the same role as the median in a classical boxplot.

### 1. Band of curves

Let us consider a *n*-sample of curves $y_1, \ldots, y_n$, and let us choose *i* curves among the sample: $y_{i_1}, \ldots, y_{i_k}$. The "band of curves" defined by $y_{i_1}, \ldots, y_{i_k}$ is the subset $B(y_{i_1}, \ldots, y_{i_k})$ defined as the set of points between the lower and the upper envelope of *k* curves, that is to say:

$$B(y_{i_1}, \ldots, y_{i_k}) = \left\{ (t, y(t)), t \in I \; ; \; \min_{i=i_1,\ldots,i_k} y_i(t) \leq y(t) \leq \max_{i=i_1,\ldots,i_k} y_i(t) \right\}$$

An illustration of this definition is given in figure 9, where a sample of curves is represented, the band is the area bounded by the two black lines.
In this article, only the case of *k*=2 curves bands is considered, which is the most useful in practice.

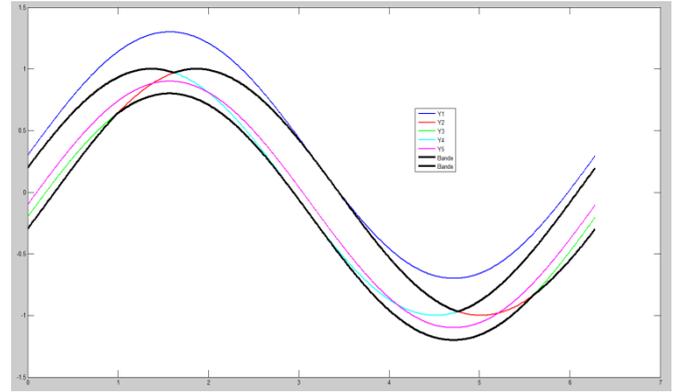

**Figure 9: Example of band of curves.**

### 2. Band depth concept

Lopez-Pintado and Romo[5] define the band depth of a curve *y*, in the case of bands of two curves by

$$\text{BD2}(y) = \binom{n}{2}^{-1} \sum_{i \neq j} 1_{y \in B(y_i, y_j)}$$

where $\binom{n}{2}$ is the number of pairs of two curves among *n*. Thus, the higher is the band depth, the more "central" is the curve position, that is to say, the more it is included in a large number of bands. If the highest band depth is reached by two different curves, the most central curve is the average of these curves where the maximal value is reached.

Figure 10 shows a sample from Sun and Genton[7] to illustrate this concept. The sample is composed of 4 curves, from which we can therefore form 6 bands of 2 curves. The shaded area corresponds to the band $B(y_1, y_3)$, the curves $y_1, y_2, y_3$ are in this band, but $y_4$ is not. We can easily calculate that $\text{BD}(y_1) = 3/6$, $\text{BD}(y_2) = 5/6$, $\text{BD}(y_3) = 3/6$ and $\text{BD}(y_4) = 3/6$. The curve $y_2$ is thus the most central (what will be called later the "central curve").

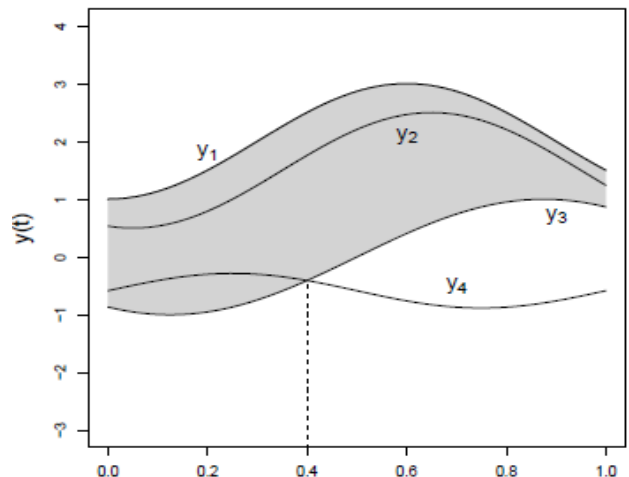

**Figure 10: Illustration of band depth. Source: Sun and Genton[7].**

## 3. Functionnal boxplot based on band depth

The functional boxplot introduced by Sun and Genton[7] is a generalization of the usual boxplot based on band depth. The principle of this type of graph is shown in figure 11, applied to the BEMUSE study-case. The dark grey area represents the central region, defined as the envelope of α- proportion of the deepest curves ($0 \leq \alpha \leq 1$). The default value α =0.5 is provided by Sun and Genton[7]; region is then written

$$C_{50\%} = B(y_{[1]}, \ldots, y_{[\frac{n}{2}]})$$

where $(y_{[1]}, \ldots, y_{[n]})$ is as previously the ranked sample by decreasing band depth. The central region is equivalent to the box in the classical boxplot. It provides a visual representation of the extent of 50% of the curves. Moreover, within this zone, the black curve represents the central (or median) curve.

Sun and Genton[7] propose to consider as outlier any curve that is not completely within a region (not shown) obtained by "increasing" the central region, down and up, by an amount at each point proportional to the height of the band. The proportionality factor (sometimes called "expansion factor" in the following) is equal to 1.5 by default, by analogy with the usual boxplot. The envelope of all curves that are not outliers is shown in the graph (light grey line). The colored curves correspond to the detected outliers.

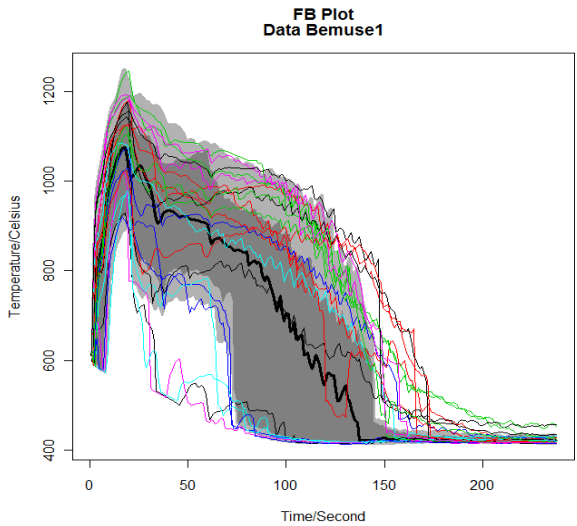

**Figure 11: Functionnal boxplot of BEMUSE study-case.**

## IV. Results on the BEMUSE study-case

The considered data set consists of 100 discretized curves of 237 sampling points (figure 1). In order not to be disturbed by the stationary regime (end of transient), which is less interesting for industrial application, we have considered only the first 150 points of each curve. The graphical representation of the bivariate space (see figure 2) seems rather unimodal, and the variance explained by the dimension reduction is quite high (85%).

Figures 4, 7 and 11 show that the confidence intervals are very similar from a method to another. We see also that there are small differences between the dark gray area and the light gray area. It means that most of the curves are in the dark gray envelope. From its shape, we can observe the small uncertainty zone at the beginning of the transient and the large increase of uncertainty during the decrease phase of the temperature.

From Figures 4, 7 and 11, we see also that outliers are quite similar among the three methods. The band depth method gives more outliers but this strongly depends on the expansion factor tuning. From Figures 4 and 7, one can guess two kinds of outliers: amplitude outliers down (visible between t = 30s and t = 100s), and amplitude outliers up (t> 150s). In order to exploit these results, a fine analysis of the combination of CATHARE2 input parameters leading to these "abnormal" results have to be made. This analysis is not realized in this work.

## V. Conclusion

In uncertainty studies, when analyzing a large number of results which are in a functional form (as time dependent curves), we are faced to difficult visualization problems/ In this paper, we have provided some methods in order to answer to three questions asked in introduction when dealing with a large number of one-dimensional curves:
1. What is the average curve?
2. Can we define some confidence interval curves containing most of the curves?
3. Can we detect some abnormal curves, in the sense of a strong difference from the majority of the curves?

The function boxplot and bagplot tools allows to answer to these three questions: the median curve for question 1, the gray areas for question 2 and the so-called outlier curves for question 3.

We have also shown that visualization tools can be helpful for thermal-hydraulical transient selection: from a large number of curves, detecting which transients have a particular shape is not obvious. This question is particularly crucial in a sensitivity analysis approach, where this kind of tools could be coupled with other graphs (as cobweb plot): when other (scalar or functional) random variables are studied, it is important to have powerful visual ranking tool to show how influent a variable or group of variables is on the output quantity of interest. Future works will develop some links between curve band depth and sensitivity analysis objectives.

## Acknowledgment

All the statistical parts of this work have been performed within the R environment, by using the packages "rainbow" and "fda". We are grateful to Arnaud Phalipaud, Dorian Deneuville and Zhijin Li who realized a large part of this work during their last year of engineering schools. This work has also been supervised by Emmanuel Vazquez and Julien Bect from Supelec, France. We thank Agnès de Crécy for providing the BEMUSE study-case.